\documentclass{llncs}

\usepackage{amsmath,bm}
\usepackage{amsfonts}
\usepackage{graphicx}
\graphicspath{{figures/}}

\begin{document}

\title{Chosen-Plaintext Cryptanalysis of a Clipped-Neural-Network-Based Chaotic
Cipher%
\thanks{This paper has been published in \textit{Advances in Neural Networks ¨C ISNN 2005:
Second International Symposium on Neural Networks, Chongqing, China, May 30 - June 1, 2005,
Proceedings, Part II} (ISNN 2005), \textit{Lecture Notes in Computer Science}, vol. 3497, pp. 630-636.}}

\author{Chengqing Li\inst{1}, Shujun Li\inst{2a}\thanks{The corresponding author, personal
web site: \url{http://www.hooklee.com}.}, Dan Zhang\inst{3} and
Guanrong Chen\inst{2b}}

\institute{Department of Mathematics, Zhejiang University,
Hangzhou 310027, China\\
\email{swiftsheep@hotmail.com} \and
Department of Electronic Engineering, City University of Hong
Kong, Kowloon, Hong Kong, China\\
\email{hooklee@mail.com}\inst{2a},
\email{eegchen@cityu.edu.hk}\inst{2b} \and College of Computer
Science, Zhejiang University, Hangzhou 310027, China\\
\email{zhangdan@etang.com}}

\maketitle

\begin{abstract}
In ISNN'04, a novel symmetric cipher was proposed, by combining a
chaotic signal and a clipped neural network (CNN) for encryption.
The present paper analyzes the security of this chaotic cipher
against chosen-plaintext attacks, and points out that this cipher
can be broken by a chosen-plaintext attack. Experimental analyses
are given to support the feasibility of the proposed attack.
\end{abstract}

\setcounter{footnote}{0}

\section{Introduction}

Since the 1990s, the study of using chaotic systems to design new
ciphers has become intensive \cite{ShujunLi:Dissertation2003}. In
particular, the idea of combining chaos and neural networks has
been developed \cite{Yen-Guo:CNN:VLSI-CAD99},
\cite{Su-Lin-Yen:CNN:APCCAS2000}, \cite{Yen-Guo:CNN:PRIA2002},
\cite{Zhou:CNN:ISNN2004} and has been adopted for image and video
encryption \cite{Lian:CNN:ISNN2004}, \cite{Lian:CNN:ICONIP2004}.
In our recent work \cite{Li:AttackingCNN2004}, it has been shown
that the chaotic ciphers designed in
\cite{Yen-Guo:CNN:VLSI-CAD99}, \cite{Su-Lin-Yen:CNN:APCCAS2000},
\cite{Yen-Guo:CNN:PRIA2002}, \cite{Lian:CNN:ISNN2004},
\cite{Lian:CNN:ICONIP2004} are not sufficiently secure from a
cryptographical point of view.

This paper focuses on the security of a
clipped-neural-network-based chaotic cipher proposed in ISNN'04
\cite{Zhou:CNN:ISNN2004}. This chaotic cipher employs a chaotic
pseudo-random signal and the output of a 8-cell clipped neural
network to mask the plaintext, along with modulus additions and
XOR operations. Also, the evolution of the neural network is
controlled by the chaotic signal. With such a complicated
combination, it was hoped that the chaotic cipher can resist
chosen-plaintext attacks. Unfortunately, our analysis shows that
it is still not secure against chosen-plaintext attacks. By
choosing only two plaintexts, an attacker can derive an equivalent
key to break the cipher. This paper reports our analyses and
simulation results.

The rest of the paper is organized as follows. Section
\ref{section:CNNC} is a brief introduction to the chaotic cipher
under study. The proposed chosen-plaintext attack is described in
detail in Sec.~\ref{section:cryptanalysis}, with some experimental
results. The last section concludes the paper.

\section{The CNN-Based Chaotic Cipher}
\label{section:CNNC}

First, the CNN employed in the chaotic cipher is introduced. The
neural network contains 8 neural cells, denoted by
$S_0,\cdots,S_7\in\{1,-1\}$, and each cell is connected with other
cells via eight synaptic weights $w_{ij}\in\{1,0,-1\}$, among
which only three are non-zeros. The synaptic weights between two
connected cells are identical: $\forall\; i,j=0\sim 7$,
$w_{ij}=w_{ji}$. The neural network evolves according to the
following rule: $\forall\; i=0\sim 7$,
\begin{equation}
f(S_i)=\mathrm{sign}\left(\widetilde{S}_i\right)=\begin{cases}
1, & \widetilde{S}_i>0\enspace,\\
-1, & \widetilde{S}_i<0\enspace,
\end{cases}
\end{equation}
where $\widetilde{S}_i=\sum\nolimits_{j=0}^7 w_{ij}S_j$. Note that
$\widetilde{S}_i\neq 0$ holds at all times.

Now, let us see how the chaotic cipher works with the above CNN.
Without loss of generality, assume that $f=\{f(i)\}_{i=0}^{N-1}$
is the plaintext signal, where $f(i)$ denotes the $i$-th
plain-byte and $N$ is the plaintext size in byte. Accordingly,
denote the ciphertext by $f'=\{f'(i)\}_{i=0}^{N-1}$, where $f'(i)$
is a double-precision floating-point number corresponding to the
plain-byte $f(i)$. The encryption procedure can be briefly
depicted as follows\footnote{Note that some original notations
used in \cite{Zhou:CNN:ISNN2004} have been changed in order to
provide a better description.}.

\begin{itemize}
\item \textit{The secret key} includes the initial states of the 8
neural cells in the CNN, $S_0(0),\cdots,S_7(0)$, the initial
condition $x(0)$, and the control parameter $r$ of the following
chaotic tent map:
\begin{equation} T(x)=
\begin{cases}
rx, & 0<x\leq 0.5\enspace,\\
r(1-x), &  0.5<x<1\enspace,
\end{cases}
\end{equation}
where $r$ should be very close to 2 to ensure the chaoticity of
the tent map.

\item \textit{The initial procedure}: 1) in double-precision
floating-point arithmetic, run the tent map from $x(0)$ for 128
times before the encryption starts; 2) run the CNN for $128/8=16$
times (under the control of the tent map, as discussed below in
the last step of the encryption procedure); 3) set $x(0)$ and
$S_0(0),\cdots,S_7(0)$ to be the new states of the tent map and
the CNN.

\item \textit{The encryption procedure}: for the $i$-th plain-byte
$f(i)$, perform the following steps to get the ciphertext $f'(i)$:
\begin{itemize}
\item evolve the CNN for one step to get its new states:
$S_0(i),\cdots,S_7(i)$;

\item in double-precision floating-point arithmetic, run the
chaotic tent map for 8 times to get 8 chaotic states:
$x(8i+0),\cdots,x(8i+7)$;

\item generate 8 bits by extracting the 4-th bits of the 8 chaotic
states: $b(8i+0),\cdots,b(8i+7)$, and then $\forall\; j=0\sim 7$,
set $E_j=2\cdot b(8i+j)-1$;

\item encrypt $f(i)$ as follows\footnote{In
\cite{Zhou:CNN:ISNN2004}, $x(8i+7)$ was mistaken as $x(8)$.}:
\begin{equation}\label{equation:encrypt}
f'(i)=\left(\left(\frac{f(i)\oplus B(i)}{256}+x(8i+7)\right)\bmod
1\right)\enspace,
\end{equation}
where $B(i)=\sum_{j=0}^7\left(\frac{S_j(i)+1}{2}\right)\cdot
2^{7-j}$;

\item $\forall\;i=0\sim 7$, if $S_i\neq E_i$, update all the three
non-zero weights of the $i$-th neural cell and the three mirror
weights as follows: $w_{ij}=-w_{ij}$, $w_{ji}=-w_{ji}$.
\end{itemize}

\item \textit{The decryption procedure} is similar to the above
one with the following decryption formula:
\begin{equation}\label{equation:decrypt}
f(i)=(256\cdot((f'(i)-x(8i+7))\bmod 1))\oplus B(i)\enspace.
\end{equation}
\end{itemize}

\section{The Chosen-Plaintext Attack}
\label{section:cryptanalysis}

In chosen-plaintext attacks, it is assumed that the attacker can
intentionally choose a number of plaintexts to try to break the
secret key or its equivalent
\cite{Schneier:AppliedCryptography96}. Although it was claimed
that the chaotic cipher under study can resist this kind of
attacks \cite[Sec. 4]{Zhou:CNN:ISNN2004}, our cryptanalysis shows
that such a claim is not true. By choosing two plaintexts, $f_1$
and $f_2$, satisfying $\forall\; i=0\sim N-1$,
$f_1(i)=\overline{f_2(i)}$, one can derive two masking sequences
as equivalent keys for decryption.

Before introducing the chosen-plaintext attack, three lemmas are
given, which are useful in the following discussions.
\begin{lemma}
$\forall\, a,b,c\in \mathbb{R}, c\neq 0$ and $n\in\mathbb{Z}^+$,
if $a=(b\mod c)$, one has $a\cdot n=((b\cdot n)\bmod(c\cdot
n))$.\label{Lemma1}
\end{lemma}
\begin{proof}
From $a=(b\mod c)$, one knows that $\exists\, k\in\mathbb{Z}$,
$b=c\cdot k+a$ and $0\leq a<c$. Thus, $\forall\;
n\in\mathbb{Z}^+$, $b\cdot n=c\cdot n\cdot k+a\cdot n$ and $0\leq
a\cdot n<c\cdot n$, which immediately leads to $a\cdot n=((b\cdot
n)\bmod(c\cdot n))$ and completes the proof of this lemma.\qed
\end{proof}

\begin{lemma}
$\forall\, a,b,c,n\in \mathbb{R}$ and $0\le a,b<n$, if
$c=((a-b)\mod n)$, one has $a-b\in\{c,c-n\}$.\label{Lemma2}
\end{lemma}
\begin{proof}
This lemma can be proved under two conditions. i) When $a\geq b$,
it is obvious that $((a-b)\bmod n)=a-b=c$. ii) When $a<b$,
$((a-b)\bmod n)=((n+a-b)\bmod n)$. Since $-n<a-b<0$, one has
$0<n+a-b<n$, which means that $((a-b)\bmod n)=n+a-b=c$. That is,
$a-b=c-n$. Combining the two conditions, this lemma is thus
proved.\qed
\end{proof}

\begin{lemma}
Assume that $a,b$ are both 8-bit integers. If $a=b\oplus 128$,
then $a\equiv(b+128)\pmod{256}$.\label{lemma3}
\end{lemma}
\begin{proof}
This lemma can be proved under two conditions. i) When $0\leq
a<128$: $b=a\oplus 128=a+128$, so $a\equiv(b+128)\pmod{256}$. ii)
When $128\leq a\leq 255$: $b=a\oplus 128=a-128$, so
$a\equiv(b-128)\equiv(b+128)\pmod{256}$. \qed
\end{proof}

From Lemma \ref{Lemma1}, one can rewrite the encryption formula
Eq. (\ref{equation:encrypt}) as follows:
\begin{equation}
256\cdot f'(i)=\left(\left((f(i)\oplus B(i))+256\cdot
x(8i+7)\right)\bmod 256\right)\enspace.
\end{equation}
Given two plain-bytes $f_1(i)\neq f_2(i)$ and the corresponding
cipher-blocks $f_1'(i),f_2'(i)$, one has
$256\cdot(f_1'(i)-f_2'(i))\equiv\left((f_1(i)\oplus
B(i))-(f_2(i)\oplus B(i))\right)\pmod{256}$. Without loss of
generality, assume that $f_1'(i)>f_2'(i)$ and that
$\Delta_{f_{1,2}}=256\cdot(f_1'(i)-f_2'(i))$. It is true that
$0<\Delta_{f_{1,2}}<256$. Thus, one has
\begin{equation}
\Delta_{f_{1,2}}=\left(\left((f_1(i)\oplus B(i))-(f_2(i)\oplus
B(i))\right)\bmod 256\right)\enspace.
\end{equation}
Because $f_1(i)\oplus B(i)$ and $f_2(i)\oplus B(i)$ are 8-bit
integers and $\Delta_{f_{1,2}}\neq 0$, from Lemma \ref{Lemma2},
one of the following facts is true:
\begin{subequations}
\begin{eqnarray}
\mbox{1. }(f_1(i)\oplus B(i))-(f_2(i)\oplus B(i)) & = &
\Delta_{f_{1,2}}\in\{1, \cdots, 255\}\enspace;\label{equation:diff1}\\
\mbox{2. }(f_2(i)\oplus B(i))-(f_1(i)\oplus B(i)) & = &
\left(256-\Delta_{f_{1,2}}\right)\in\{1, \cdots,
255\}\enspace.\label{equation:diff2}
\end{eqnarray}
\end{subequations}
For the above two equations, when
$f_1(i)=\overline{f_2(i)}$ is satisfied, two possible values of
$B(i)$ can be uniquely derived according to the following theorem.
\begin{theorem}
Assume that $a,b,c,x$ are all 8-bit integers, and $c>0$. If
$a=\bar{b}$, then the equation $(a\oplus x)-(b\oplus x)=c$ has an
unique solution $x=a\oplus (1,c_7,\cdots,c_1)_2$, where
$c=(c_7,\cdots,c_0)_2=\sum_{i=0}^7c_i\cdot
2^i$.\label{theorem1}\end{theorem}
\begin{proof}
Since $a=\bar{b}$, one has $b\oplus x=\overline{a\oplus x}$. Thus,
by substituting $y=a\oplus x$ and $\bar{y}=\overline{a\oplus
x}=b\oplus x$ into $(a\oplus x)-(b\oplus x)=c$, one can get
$y-\bar{y}=c$, which is equivalent to $y=\bar{y}+c$. Let
$y=\sum_{i=0}^7y_i\cdot 2^i$, and consider the following three
conditions, respectively.

1) When $i=0$, from $y_0\equiv(\bar{y}_0+c_0)\pmod 2$, one can
immediately get $c_0=1$. Note the following two facts: i) when
$y_0=0$, $\bar{y}_0+c_0=2$, a carry bit is generated for the next
bit, so $y_1\equiv(\bar{y}_1+c_1+1)\pmod 2$ and $c_1=0$; ii) when
$y_0=1$, $\overline{y_0}+c_0=1$, no carry bit is generated, so
$y_1\equiv(\bar{y}_1+c_1)\pmod 2$ and $c_1=1$. Apparently, it is
always true that $y_0=c_1$. Also, a carry bit is generated if
$c_1=0$ is observed.

2) When $i=1$, if there exists a carry bit, set
$c_1'=c_1+1\in\{1,2\}$; otherwise, set $c_1'=c_1\in\{0,1\}$. From
$y_1\equiv(\bar{y}_1+c_1')\pmod 2$, one can immediately get
$c_1'=1$. Then, using the same method shown in the first
condition, one has $y_1=c_2$ and knows whether or not a carry bit
is generated for $i=2$. Repeat the above procedure for $i=2\sim
6$, one can uniquely determine that $y_i=c_{i+1}$.

3) When $i=7$, it is always true that the carry bit does not
occur, so $c_7'=1$, and $y_7\equiv 1$.

Combining the above three conditions, one can get $y=(1,c_7,\cdots
,c_1)_2$, which results in $x=a\oplus(1,c_7,\cdots,c_1)_2$. \qed
\end{proof}
Assume that the two values of $B(i)$ derived from Eqs.
(\ref{equation:diff1}) and (\ref{equation:diff2}) are $B_1(i)$ and
$B_2(i)$, respectively. The following corollary shows that the two
values have a deterministic relation: $B_2(i)=B_1(i)\oplus 128$.
\begin{corollary}
Assume that $a,b,c,x$ are all 8-bit integers, $a=\bar{b}$ and
$c>0$. Given two equations, $(a\oplus x)-(b\oplus x)=c$ and
$(b\oplus x')-(a\oplus x')=c'$, if $c'=256-c$, then $x'=x\oplus
128$.
\end{corollary}
\begin{proof}
Since $c+\bar{c}=255$, one has $c'=256-c=\bar{c}+1$. Let
$c=\sum_{i=0}^7c_i\cdot 2^i$, and observe the first condition of
the proof of Theorem \ref{theorem1}. One can see that $c_0=1$, so
$c_0'=\bar{c}_0+1=1$. Since there is no carry bit, one can deduce
that $\forall\,i=1\sim 7$, $c_i'=\bar{c}_i$. Applying Theorem
\ref{theorem1} for $(a\oplus x)-(b\oplus x)=c$, one can uniquely
get $x=a\oplus(1,c_7,\cdots,c_1)_2$. Then, applying Theorem
\ref{theorem1} for $(b\oplus x')-(a\oplus x')=c'$, one has
$x'=b\oplus(1, c_7', \cdots, c_1')_2=\bar{a}\oplus(1, \bar{c}_7,
\cdots, \bar{c}_1)_2=(a_7, \bar{a}_6\oplus \bar{c}_7,\cdots,
\bar{a}_0\oplus\bar{c}_1)_2=(a_7,a_6\oplus c_7,\cdots, a_0\oplus
c_1)_2=a\oplus (1, c_7,\cdots, c_1)_2\oplus(1,
0,\cdots,0)_2=x\oplus 128$. Thus, this corollary is proved. \qed
\end{proof}

For any one of the two candidate values of $B(i)$, one can further
get an equivalent chaotic state $\hat{x}(8i+7)$ from $B(i)$,
$f(i)$ and $f'(i)$ as follows:
\begin{equation}
\hat{x}(8i+7)=256\cdot f'(i)-(f(i)\oplus B(i))\equiv 256\cdot
x(8i+7)\pmod{256}\enspace.\label{equation:xhat}
\end{equation}
With $B(i)$ and $\hat{x}(8i+7)$, the encryption formula Eq.
(\ref{equation:encrypt}) becomes
\begin{equation}
f'(i)=\frac{\left((f(i)\oplus B(i))+\hat{x}(8i+7)\right)\bmod
256}{256}\enspace,\label{equation:encryption2}
\end{equation}
and the decryption formula Eq.~(\ref{equation:decrypt}) becomes
\begin{equation}
f(i)=\left(\left(256\cdot f'(i)-\hat{x}(8i+7)\right)\bmod
256\right)\oplus B(i)\enspace.\label{equation:decryption2}
\end{equation}

Assume that $\hat{x}_1(8i+7)$ and $\hat{x}_2(8i+7)$ are calculated
by Eq. (\ref{equation:xhat}), from $B_1(i)$ and $B_2(i)$,
respectively. Then, we have the following proposition.
\begin{proposition}
$\left(B_1(i),\hat{x}_1(8i+7)\right)$ and
$\left(B_2(i),\hat{x}_2(8i+7)\right)$ are equivalent for the above
encryption procedure Eq.~(\ref{equation:encryption2}), though only
one corresponds to the correct value generated from the secret
key. That is,
\[
\left((f(i)\oplus
B_1(i))+\hat{x}_1(8i+7)\right)\equiv\left((f(i)\oplus
B_2(i))+\hat{x}_2(8i+7)\right)\pmod{256}\enspace.
\]
\end{proposition}
\begin{proof}
From $B_1(i)=B_2(i)\oplus 128$, one has $f(i)\oplus
B_1(i)=(f(i)\oplus B_2(i)\oplus 128)$. Then, following Lemma
\ref{lemma3}, it is true that $(f(i)\oplus
B_1(i))\equiv((f(i)\oplus B_2(i))+128)\pmod{256}$. As a result,
$\hat{x}_1(8i+7)=(256\cdot f'(i)-(f(i)\oplus
B_1(i)))\equiv(256\cdot f'(i)-((f(i)\oplus
B_2(i))-128))\pmod{256}\equiv(\hat{x}_2(8i+7)+128)\pmod{256}$,
which immediately leads to the following fact: $\left((f(i)\oplus
B_1(i))+\hat{x}_1(8i+7)\right)\equiv\left((f(i)\oplus
B_2(i))+\hat{x}_2(8i+7)\right)\pmod{256}$. Thus, this proposition
is proved. \qed
\end{proof}
Considering the symmetry of the encryption and decryption
procedures, the above proposition immediately leads to a
conclusion that $\left(B_1(i),\hat{x}_1(8i+7)\right)$ and
$\left(B_2(i),\hat{x}_2(8i+7)\right)$ are also equivalent for the
decryption procedure Eq.~(\ref{equation:decryption2}).

From the above analyses, with two chosen plaintexts $f_1$ and
$f_2=\bar{f}_1$, one can get the following two sequences:
$\{B_1(i),\hat{x}_1(8i+7)\}_{i=0}^{N-1}$ and
$\{B_2(i),\hat{x}_2(8i+7)\}_{i=0}^{N-1}$. Given a ciphertext
$f'=\{f'(i)\}_{i=0}^{N-1}$, $\forall\; i=0\sim N-1$, one can use
either $(B_1(i),\hat{x}_1(8i+7))$ or $(B_2(i),\hat{x}_2(8i+7))$ as
an equivalent of the secret key to decrypt the $i$-th plain-byte
$f(i)$, following Eq. (\ref{equation:decryption2}). This means
that the chaotic cipher under study is not sufficiently secure
against the chosen-plaintext attack.

To demonstrate the feasibility of the proposed attack, some
experiments have been performed for image encryption, with secret
key $r=1.99$, $x(0)=0.41$ and
$[S_0(0),\cdots,S_7(0)]=[1,-1,1,-1,1,-1,1,-1]$. One plain-image
``Lenna'' of size $256\times 256$ is chosen as $f_1$ and another
plain-image is manually generated as follows: $f_2=\bar{f}_1$. The
two plain-images and their cipher-images are shown in
Fig.~\ref{figure:ChosenImage}. With the two chosen plain-images,
two sequences, $\{B_1(i),\hat{x}_1(8i+7)\}_{i=0}^{256\times
256-1}$ and $\{B_2(i),\hat{x}_2(8i+7)\}_{i=0}^{256\times 256-1}$,
are generated by using the above-mentioned algorithm. The first
ten elements of the two sequences are given in
Table~\ref{table:EquivalentKey}. $\forall\; i=0\sim(256\times
256-1)$, either $(B_1(i),\hat{x}_1(8i+7))$ or
$(B_2(i),\hat{x}_2(8i+7))$ can be used to recover the plain-byte
$f(i)$. As a result, the whole plain-image (``Peppers" in this
test) can be recovered as shown in Fig.~\ref{figure:ChosenImage}f.

\begin{table}[!htb]
\centering \caption{The first ten elements of
$\{B_1(i),\hat{x}_1(8i+7)\}_{i=0}^{256\times 256-1}$ and
$\{B_2(i),\hat{x}_2(8i+7)\}_{i=0}^{256\times 256-1}$}
\label{table:EquivalentKey}
\begin{tabular}{c|*{9}{c|}c}
\hline $i$ & 0& 1 & 2 & 3 & 4 & 5 & 6 & 7 & 8 & 9\\
\hline $B_1(i)$ & 146 & 231 & 54 & 202 & 59 & 243 & 166 & 173 & 233 & 82\\
\hline $B_2(i)$ & 18  & 103 & 182 & 74 & 187 & 115 & 38 & 45 & 105 & 210\\
\hline\hline $\hat{x}_1(8i+7)$ & 242.40 & 38.63 & 242.62 & 222.09
& 81.03 & 214.73 & 240.91 & 203.59 & 138.20 & 9.33\\
\hline $\hat{x}_2(8i+7)$ & 114.40 & 166.63 & 114.62 & 94.09 & 209.03 & 86.73 & 112.91 & 75.59 & 10.20 & 137.33\\
\hline
\end{tabular}
\end{table}

\newlength\mpwidth
\setlength\mpwidth{0.32\textwidth}
\newlength\figwidth
\setlength\figwidth{0.29\textwidth}

\begin{figure}[!htb]
\centering
\begin{minipage}[t]{\mpwidth}
\raggedright
\includegraphics[width=\figwidth]{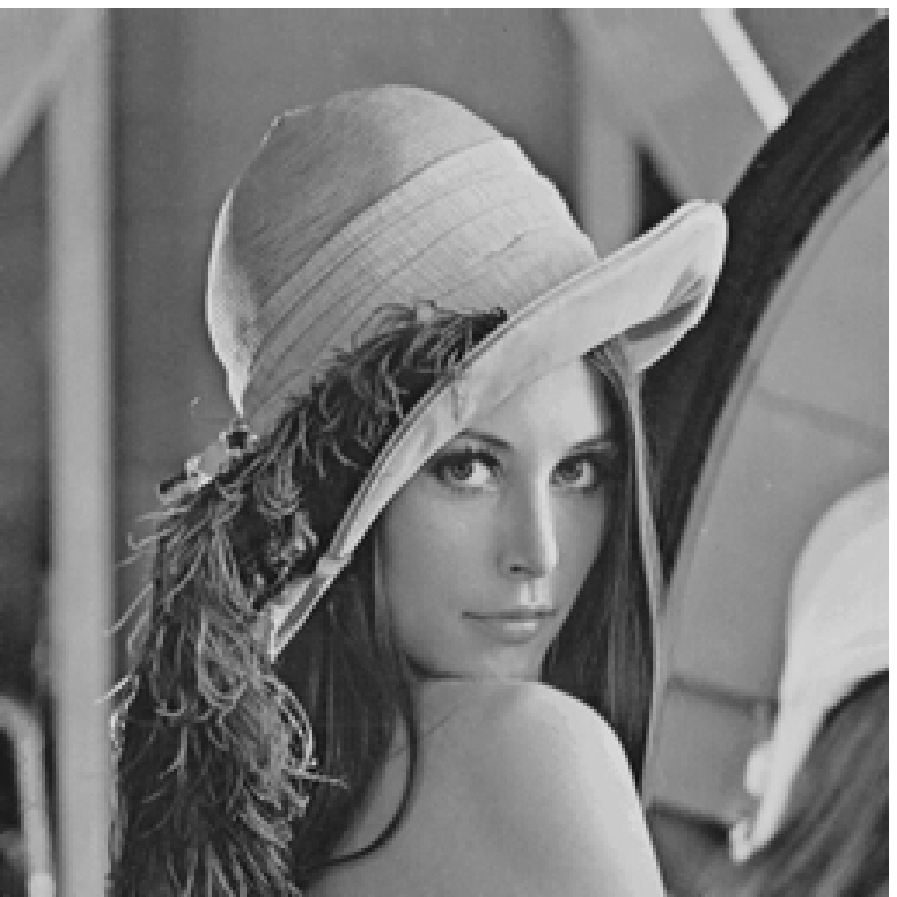}\\
a) Chosen plain-image $f_1$
\end{minipage}
\begin{minipage}[t]{\mpwidth}
\centering
\includegraphics[width=\figwidth]{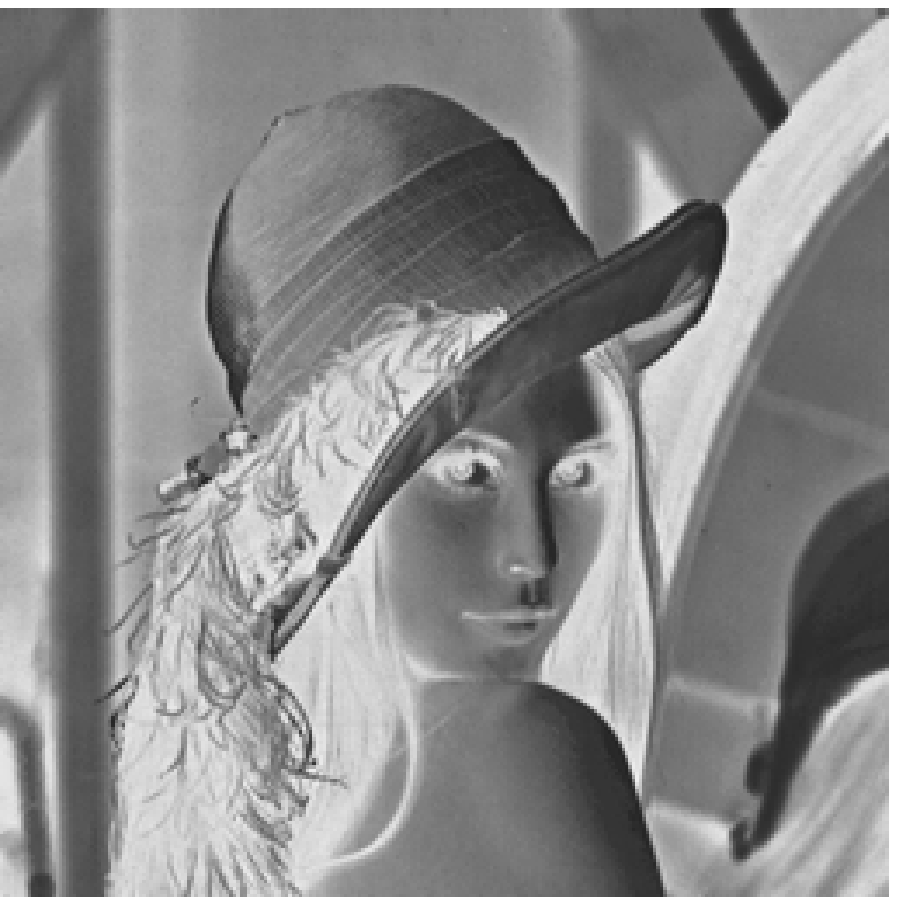}\\
c) Chosen plain-image $f_2$
\end{minipage}
\begin{minipage}[t]{\mpwidth}
\raggedright
\includegraphics[width=\figwidth]{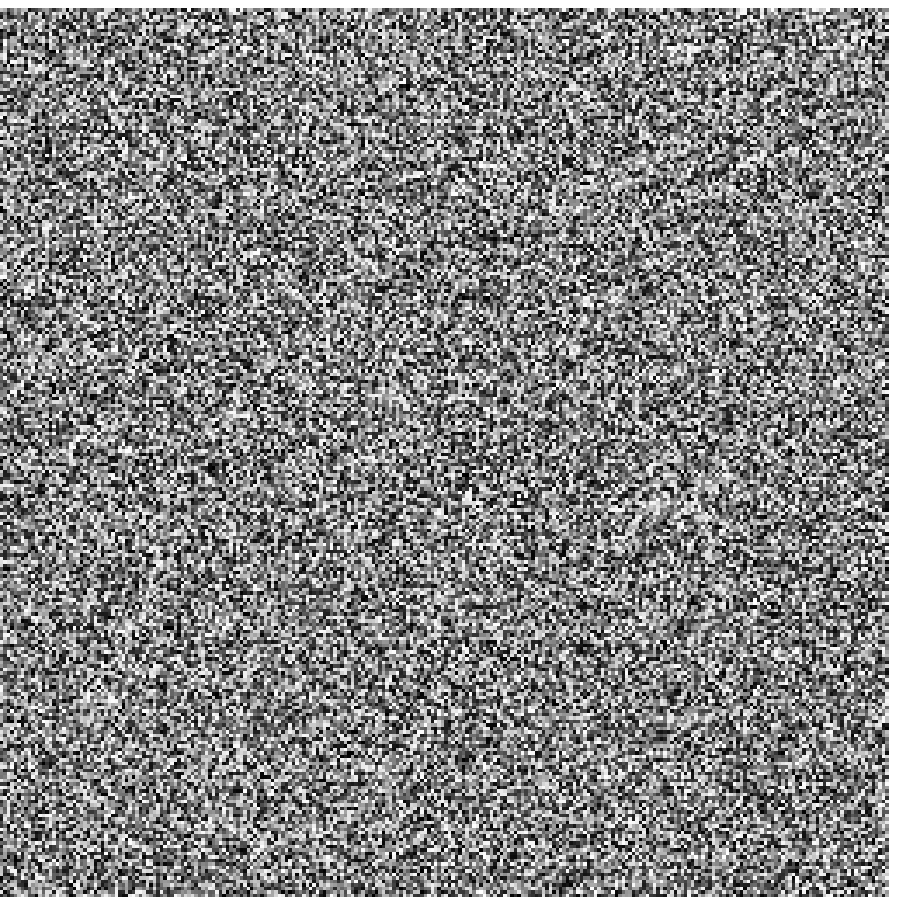}\\
e) A cipher-image $f_3'$
\end{minipage}
\begin{minipage}[t]{\mpwidth}
\raggedright
\includegraphics[width=\figwidth]{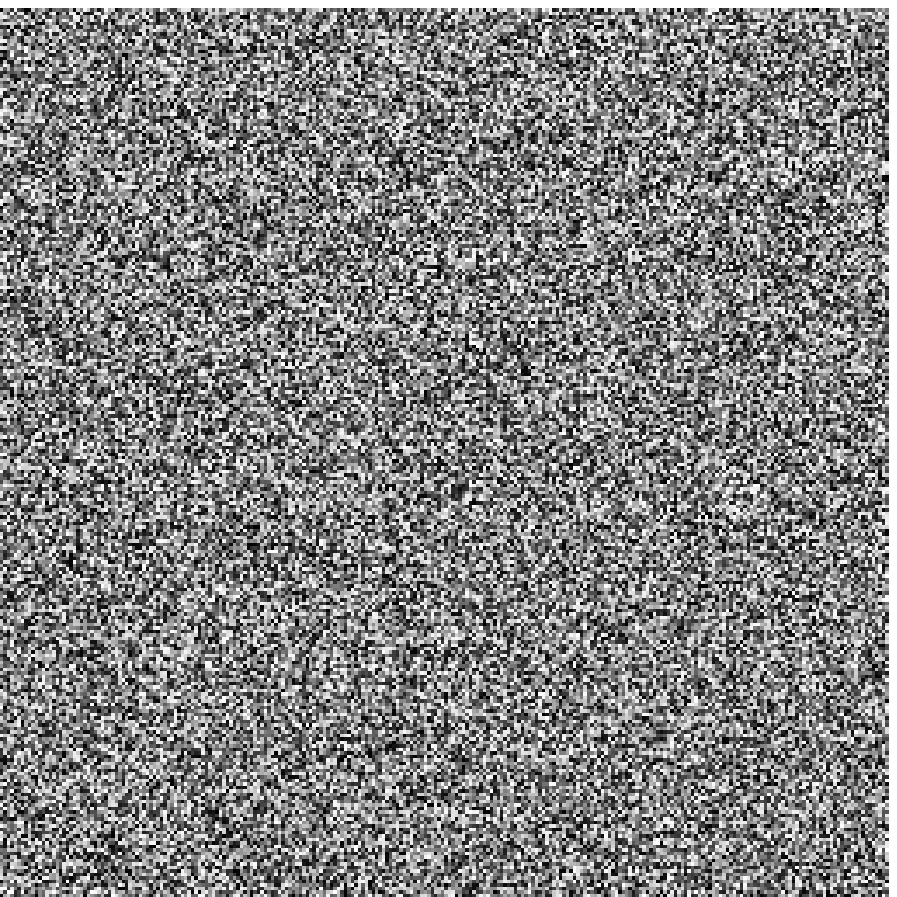}\\
b) Cipher-image $f_1'$
\end{minipage}
\begin{minipage}[t]{\mpwidth}
\raggedright
\includegraphics[width=\figwidth]{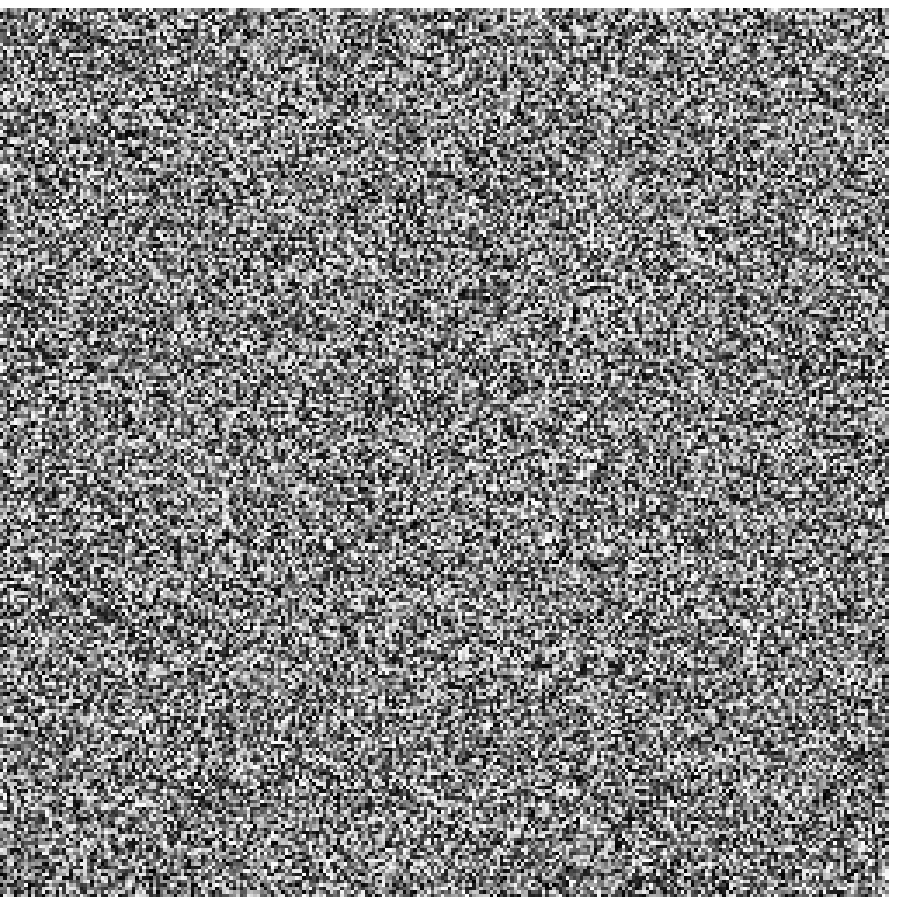}\\
d) Cipher-image $f_2'$
\end{minipage}
\begin{minipage}[t]{\mpwidth}
\raggedright
\includegraphics[width=\figwidth]{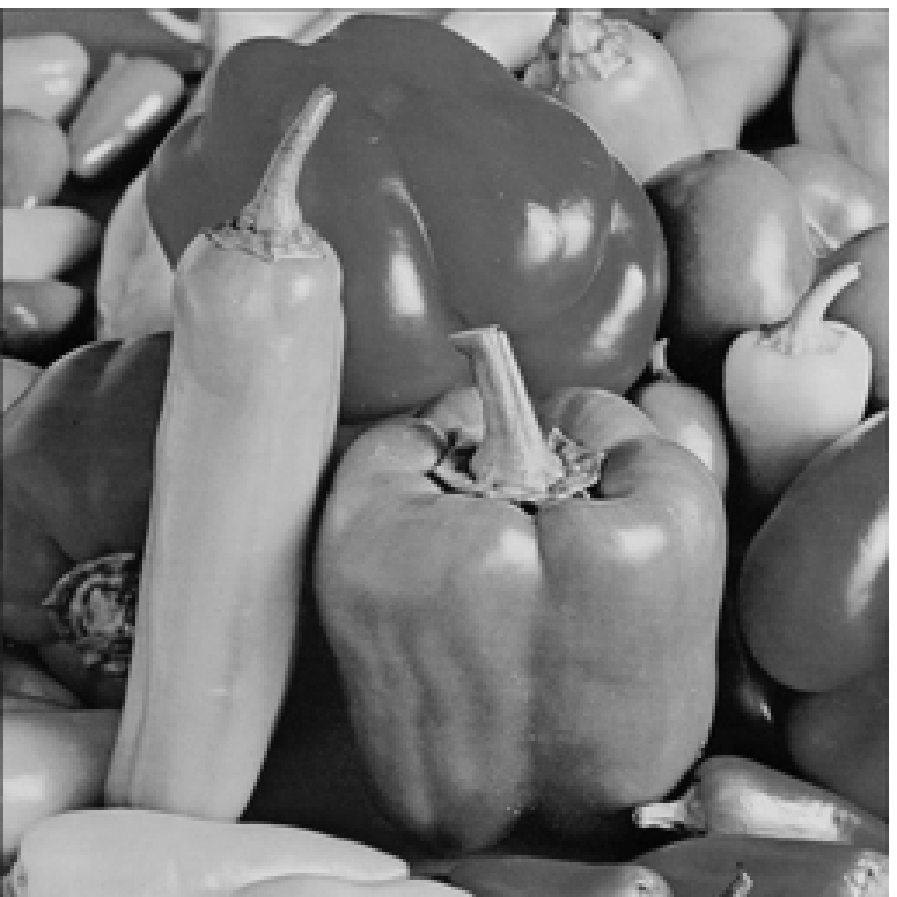}\\
f) Recovered image $f_3$
\end{minipage}
\caption{The proposed chosen-plaintext
attack}\label{figure:ChosenImage}
\end{figure}

\section{Conclusion}

In this paper, the security of a chaotic cipher based on clipped
neural network has been analyzed in detail. It is found that the
scheme can be effectively broken with only two chosen
plain-images. Both theoretical and experimental analyses have been
given to support the proposed attack. Therefore, this scheme is
not suggested for applications that requires a high level of
security.

\textbf{Acknowledgements.} This research was supported by the
National Natural Science Foundation, China, under grant no.
60202002, and by the Applied R\&D Centers of the City University
of Hong Kong under grants no. 9410011 and no. 9620004.

\bibliographystyle{splncs}
\bibliography{ISNN}
\end{document}